\documentclass[prl,aps,floatfix,twocolumn]{revtex4}
\usepackage{rotate,bm,graphicx}
\pdfoutput=1
\newcommand{\be}{\begin{equation}}
\newcommand{\ee}{\end{equation}}
\newcommand{\ber}{\begin{eqnarray}}
\newcommand{\eer}{\end{eqnarray}}

\begin{document}
\title{Ultrafast dynamics of a magnetic antivortex - Micromagnetic simulations}
\author{Sebastian Gliga}
\affiliation{Institut f\"ur Festk\"orperforschung IFF-9 "Elektronische
   Eigenschaften", Forschungszentrum J\"ulich GmbH, D-52425 J\"ulich,
   Germany}
\author{Ming Yan}
\affiliation{Institut f\"ur Festk\"orperforschung IFF-9 "Elektronische
   Eigenschaften", Forschungszentrum J\"ulich GmbH, D-52425 J\"ulich,
   Germany}
\author{Riccardo Hertel}
\affiliation{Institut f\"ur Festk\"orperforschung IFF-9 "Elektronische
   Eigenschaften", Forschungszentrum J\"ulich GmbH, D-52425 J\"ulich,
   Germany}
\author{Claus M.~Schneider}
\affiliation{Institut f\"ur Festk\"orperforschung IFF-9 "Elektronische
   Eigenschaften", Forschungszentrum J\"ulich GmbH, D-52425 J\"ulich,
   Germany}
\date{10 January 2008}

\begin{abstract}
The antivortex is a 
fundamental magnetization structure which is the topological 
counterpart of the well-known magnetic
vortex. We study here the ultrafast dynamic behavior of an isolated
antivortex in a patterned Permalloy thin-film element.
Using micromagnetic simulations we predict that the antivortex
response to an ultrashort external field pulse is characterized by the
production of a new antivortex as well as of a temporary vortex,
followed by an annihilation process. These processes are complementary
to the recently reported response of a vortex and, like for the
vortex, lead to the reversal of the orientation of the antivortex core
region. In addition to its fundamental interest, this dynamic magnetization
process could be used for the generation and propagation of spin waves
for novel logical circuits. 
\end{abstract}

\maketitle
\begin{figure*}[ht]
\centerline{\includegraphics[width=\linewidth]{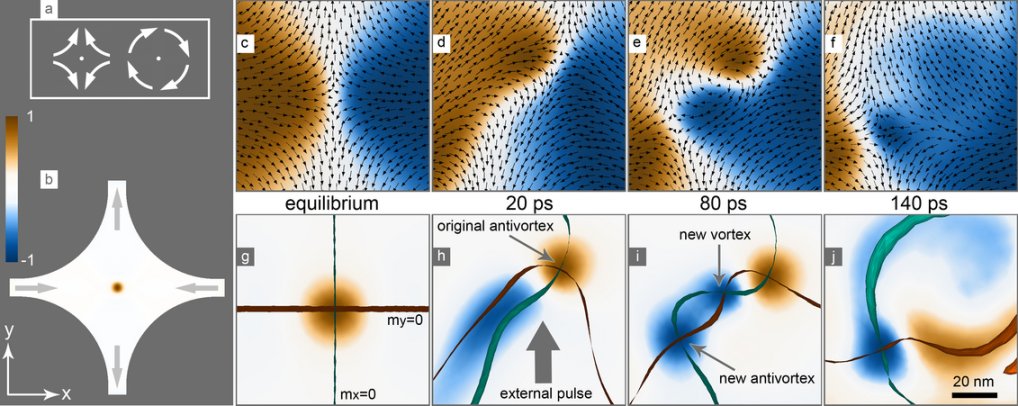}}
\caption{\label{plate}
    (color) {\it a}: Schematic of an antivortex (left) compared with a vortex
    (right). The arrows represent the in-plane magnetization. 
    {\it b}: Modeled concave-shaped sample. 
   The magnetization direction in the branches is schematically
   represented by the grey arrows. The brown area at the center of the
   structure represents the perpendicular component of the
   magnetization $m_z$, showing the antivortex core. {\it c-j}:
   Distortion of the antivortex structure induced by an in-plane
   field pulse. The frames are
   snapshots at different times relative to the 
   applied pulse's maximum. The arrows in the top row {\it (c-f)} represent
   the in-plane magnetization, while the colors represent the
   $x$-component of the magnetization ($m_x$). In the bottom row {\it
   (g-j)} the green and red ``ribbons'' are the $m_x=0$ and
   $m_y=0$ isosurfaces, respectively.  The colors represent $m_z$. 
   } 
\end{figure*}

Extended ferromagnetic films often display complex magnetization
patterns with a rich variety of features \cite{Hubert98}. 
This complexity can be reduced by decomposing the magnetic patterns
into a few elementary magnetization structures, such as 
domains, domain walls or vortices. The dynamic properties of such
fundamental structures have been investigated thoroughly over the last
years by isolating them in patterned elements~\cite{Hillebrands06}. In  
particular, the magnetic vortex has 
attracted much
interest~\cite{Wachowiak02,Novosad02,Park03b,Choe04,Stoll_n06,Hertel07,Xiao06}.
An equally fundamental, yet much less studied magnetic structure is the
antivortex, the topological counterpart of the vortex. In the complex
structures occurring in extended soft-magnetic films, antivortices can
be found almost as frequently as ordinary vortices: they occur in
cross-tie domain walls, where they are enclosed by two adjacent vortex
structures~\cite{Huber58}. While the in-plane magnetization
distribution of an antivortex is very different from that of a vortex
(see Fig.~\ref{plate}{\it a}), it contains, like the vortex, a tiny
core \cite{Wachowiak02} at its center in which the magnetization
points perpendicular to the 
plane. Moreover, the two structures are related by underlying
topological properties: In both cases, the local  magnetization
rotates by 360$^\circ$ on a closed loop around the core. The 
structures differ by their opposite sense of rotation along such a loop, which
is quantified by the winding number $w$ ($w=-1$ for the
antivortex, $w=+1$ for the vortex)\cite{bubbles79,Hertel06}.
The winding number has been predicted to have a direct impact on the
magnetization dynamics \cite{bubbles79,Thiele73}. However, not much is  
known to date about the dynamic properties of {\em anti}vortices, even though
several other fundamental magnetization structures have been analyzed
in patterned thin-film elements~\cite{Hillebrands06}.   
Studies with high spatial and temporal resolution have demonstrated
that domains, domain walls and vortices exhibit different excitation
spectra \cite{Raabe05,Park03b}. Such investigations have mostly focussed on 
small perturbations and on reversible changes in these structures
produced by an external field \cite{Wang07,Fassbender07}. However, in
a recent study on cross-tie walls, Neudert {\it et
  al.}~\cite{Neudert07}  have reported the creation of new cross-ties,
{\it i.e.}, of vortex-antivortex pairs, in 
response to high-frequency magnetic fields. Also the field-pulse induced 
  reversal of vortex cores has been found to be characterized by the
  creation of a vortex-antivortex
  pair~\cite{Hertel07,Stoll_n06,Xiao06}. During this core switch, a  
transient cross-tie structure dissolves within a few tens of ps via an
annihilation process~\cite{Lee05,Hertel06}. 
Using fully three-dimensional micromagnetic simulations we study here  
the ultrafast dynamic response of a single magnetic antivortex in response to
external field pulses a few tens of ps long. We find that a rapid series of
pair creation and annihilation processes occurs, which is complementary to the
one observed in a vortex \cite{Hertel07} and which leads to the reversal of the
antivortex core. This represents a new fundamental process in magnetism
on the nanoscale. The spin-wave bursts resulting from the annihilation
process make the antivortex structure a possible source for
spin-waves to be used in novel logical circuits~\cite{Hertel04}.

Unlike vortices, it is rather difficult to isolate an antivortex, {\em i.e.},
to prepare a nanostructure that contains only a single antivortex.
This has been achieved by Shigeto {\em et 
  al.}~using thin-film elements consisting of four connected rings
  \cite{Shigeto02}, where an antivortex was observed at the ring junction.
Here we use a sample with a shape corresponding to the connecting
part of four adjacent rings to stabilize an isolated antivortex (shown in  
Fig.~\ref{plate}{\it b}). The sample is constructed from four
circular segments with 200~nm radius. The four short stretching strips
are 40~nm wide, and the thickness is 20~nm. The in-plane shape
anisotropy of this particular geometry can sustain an antivortex due
to the tendency of the magnetization to align with the sample
boundaries. The simulations have been performed with the micromagnetic 
finite-element code that we have used previously, e.g., in
Ref.~\cite{Hertel07}.  
The antivortex structure was obtained by selecting a suitable
initial magnetic configuration and letting the structure relax
\footnote{The
  antivortex structure is not the only stable magnetization 
  state here. Using dynamic relaxation methods,
  we have equally observed the formation of a metastable vortex
  state and a state containing two domain walls. 
}. 
The material parameters used in the simulations are those of
  Permalloy: $A=13$\,pJ/m  (exchange constant), $\mu_0M_{\rm
  s}=1.0$\,T (with $M_s$ the saturation magnetization), 
$K$=0 (anisotropy constant) and $K_s=10^{-4}$\,J/m$^2$ (surface
  anisotropy) \cite{Rantschler05}. About 200,000 
  tetrahedral elements were used for the discretization of the sample.
  This corresponds to a cell size of about 4\,nm$^3$. The
  magnetization dynamics was calculated using the
  Landau-Lifshitz-Gilbert equation with damping parameter
  $\alpha=0.01$. The isosurface representation introduced in
  Ref.~\cite{Hertel06} has been used to highlight the location of the
  antivortex core, that is, the region where the magnetization is
  exactly perpendicular to the sample plane ($m_z=\pm1$). 
  This situation 
  occurs where the $m_x=0$ and $m_y=0$ isosurfaces
  intersect ($\bm{m}$ is the normalized magnetization:
  $\bm{m}=\bm{M}/M_s$).  

The antivortex structure was perturbed by a short in-plane
Gaussian-shaped pulse applied in the $y$ direction. In the case shown
in Fig.~\ref{plate}, the pulse had a maximum intensity of 60~mT and a
duration of $\sigma=80$\,ps.  Fig.~\ref{plate}{\it c-j} shows the
magnetization dynamics in response to the applied field pulse. In the
top row {\it (c-f)}, the in-plane magnetic structure is shown. 
The same structure is also shown in the bottom row {\it (g-j)}, 
but the colors there represent 
the perpendicular component of the magnetization 
($m_z$) and the crossing point of the isosurfaces helps display the 
position of the antivortex core. Following the application of the
field pulse, the antivortex  is displaced from its equilibrium
position and the in-plane magnetization of the sample is distorted
{\it (d)}. In a region close to the core, this distortion leads to the
formation of a ``dip'' in which the magnetization rotates out of the
plane in the direction opposite to the antivortex core {\it (h)}. The
formation of such a dip in the case of an antivortex is 
analogous to the recently reported dip formation near a distorted
vortex core \cite{Hertel07,Novosad05,Kasai06}. Its direction is determined
by the stray field of the antivortex core~\cite{Hertel07}. 
Approximately 80\,ps after the pulse maximum, a new antivortex is
emitted from the original antivortex {\it (e)}. 
However, according to the Poincar\'e-Hopf index theorem of vector fields, the
total winding number of the magnetization in a thin-film element is a
topological invariant, and the the formation of a single antivortex
would violate its conservation. A new vortex is thus also produced,
which is located between the two antivortex structures. The newly formed
antivortex-vortex pair is made clearly visible by the two additional
intersections of the isosurfaces {\it (i)}. This strongly
inhomogeneous structure is eventually resolved through the
annihilation of the initial antivortex with the newly created vortex,
whose cores have opposite orientation \cite{Hertel06}. This results
in a sudden reduction of the local exchange energy density and leaves
behind the newly formed antivortex with oppositely magnetized core
{\it (f, j)}. 

The production of an antivortex-vortex pair from an antivortex structure is a
previously unreported micromagnetic process. The resulting
transient magnetic configuration could be called an {\it anti} cross-tie wall,
{\it i.e.}\ a single vortex enclosed by two antivortices, which has never
been observed in a stable magnetic structure. 
It is also remarkable that, in spite of the very different magnetic in-plane 
structure of vortices and antivortices, their ultrafast dynamics are
analogous, involving the creation of new magnetic structures followed by a
destruction process.
Here, the strength of the isosurface representation is obvious: it
captures the main features of the processes unfolding 
in both vortex and antivortex structures \footnote{The
  isosurface representation of the antivortex core switching is almost
  identical to the isosurface representation of the vortex core
  switching, cf.~Fig.~2 of Ref.~ \cite{Hertel07}}, demonstrating the
complementarity between them. 

\begin{figure}[!]
\centerline{\includegraphics[width=.95\linewidth]{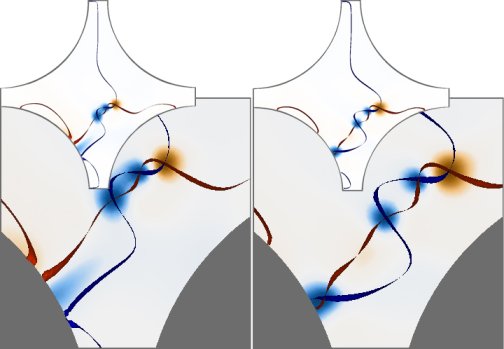}}
\caption{\label{nucleation}
(color) Vortex nucleation at the sample edge during the antivortex-initiated
pair creation process. {\it Left}: For a Gaussian pulse of 60 mT and 80 ps
duration, no edge nucleation occurs. {\it Right}: For a stronger pulse
(80 mT) of same duration, the $m_x$ and $m_y$ isosurfaces cross at
the sample boundary, indicating the nucleation of a vortex.}  
\end{figure}

\begin{figure}[!]
\centerline{\includegraphics[width=.95\linewidth]{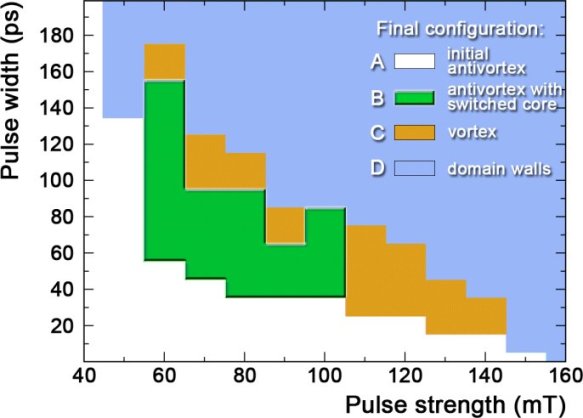}}
\caption{\label{switch_diagram}
(color online) Final magnetic configurations depending on the field pulse strength
and duration. The duration is defined as the width of the
Gaussian. The field was varied in increments of 10 ps and for every 10
mT.} 
\end{figure}

In the case shown in Fig.\ \ref{plate}, the system eventually relaxes
back to an antivortex structure. 
However, the stability of an isolated antivortex is rather weak
compared to that of a vortex. This is due to the magnetic flux, which in an
antivortex has a saddle-point configuration while a (energetically
favorable) closed flux is obtained  around an ordinary vortex. The
magnetic configuration in the sample we have used 
leads in fact to the formation of magnetic surface and volume charges,
which reduce the stability of the antivortex  \footnote{S. Gliga,
  R. Hertel and C. M. Schneider (unpublished).}. This easily
introduces the possibility of a vortex nucleating at the sample
boundaries if stronger external field pulses are applied.
Using isosurfaces, the vortex nucleation process at our sample's boundaries
can easily be described and visualized. The behavior of the 
$m_x=0$ and $m_y=0$ isosurfaces is shown in Fig.~\ref{nucleation} for two
field pulse strengths. The crossing of the isosurfaces at the boundaries
indicates that a vortex is nucleated
\footnote{
The
isosurfaces are perpendicular to the sample 
edges, which is a result of the  Brown boundary condition 
$\left.\bm{\nabla}m_i\right|_{\partial\Omega}\cdot\bm{n}=0,\,(i=x,y,z)$,
where $\bm{n}$ is the unit vector normal to the surface $\partial\Omega$.
The magnetization gradient is therefore
non-vanishing only along a direction parallel to the sample
edges. As isosurfaces and gradient lines are perpendicular to each other,
the $m_x=0$ and $m_y=0$ isosurfaces are always perpendicular to the sample
edges.}.
The introduction of a vortex in the sample affects the dynamics of the
antivortex and thus also the final equilibrium structure. 
The different possible final states are shown in
Fig.\ \ref{switch_diagram} as a function of the applied pulse's strength and
duration. It indicates that while it is rather difficult to isolate an
antivortex, it is pretty easy to dissolve it. Starting with low and short
field pulses, there is a set of parameters for which the system is
slightly excited, causing the antivortex core to rotate about its
equlibrium position (region A in Fig.~\ref{switch_diagram}). The modes excited by such low
field-pulses have been investigated in Ref.~\cite{Wang07}. When pair
production and, consequently, a core switch is induced, the 
studied system relaxes into an antivortex configuration only for a rather narrow
set of field pulse parameters (region B). For stronger or longer pulses,
vortex nucleation occurs concurrently to the antivortex
core switch, as shown in Fig.~\ref{nucleation}. The vortex can then 
migrate through the sample, leading to the expulsion of the antivortex
after a few ns (region C). In region D (for pulses above 60
mT), the pulse is sufficiently strong to induce the production of new
pairs originating from {\it both} the antivortex and the nucleated
vortex. The result is the generation of a transient state containing
multiple cross-ties. A series of annihilations
\cite{Lee04,Lee05,Hertel06} occurs within a few ten ns, leading to a a  
final magnetic configuration consisting only of a pair of head-to-head
and tail-to-tail domain walls.  

\begin{figure}[!]
\centerline{\includegraphics[width=\linewidth]{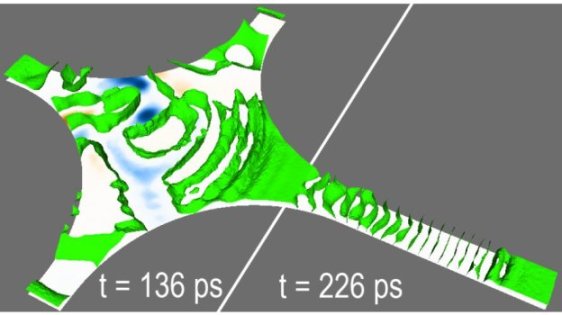}}
\caption{\label{propagation}
(color) Spin wave propagation 
at 136 ps (left) and 226 ps (right)
after the 
pulse's maximum.
The spin waves generated by the core reversal (left) shortly later
propagate smoothly into the branch (right).
The $m_z=0$ isosurfaces shown in green allow to visualize the wave
fronts. The blue spot represents the switched core ($m_z$) using the
same scale as in Fig.~\ref{plate}.} 
\end{figure}

The annihilation of a vortex and an antivortex
with opposite core magnetization 
has been shown to be connected with the emission of
spin-wave bursts in the GHz range~\cite{Lee05,Hertel06}.
It has therefore been suggested that the core switch mechanism of a
{\it vortex} could be used to inject spin waves into a strip acting as
a waveguide \cite{Choi07}. The spin waves could then be processed in
logical circuits \cite{Hertel04}. In this context, the antivortex
structure could equally be used as a spin-wave source. If its
stability can be increased, an antivortex would probably be more
suitable than a vortex for this purpose: When a disk-shaped source is
used (in the case of the {\it vortex} core switch) a
90$^\circ$N\'{e}el wall is inevitably present between the source and
the strip along which the waves propagate. Such domain walls have been
reported to have unfavorable properties concerning spin wave
propagation, mostly reflecting the spin waves \cite{Buess06}. In
contrast, as shown in Fig.~\ref{propagation}, the spin waves produced
in the  antivortex can propagate unhindered into a branch, which
naturally extends the magnetic configuration of the astroid-like
sample. All branches can be extended to connect to other astroid-like
elements. Such a network could constitute an antidot array
\cite{Heyderman07} and provide a regular lattice of antivortices.  

What does the study of antivortex dynamics lead to?
First, the complex and ultrafast modifications that an antivortex can
undergo by applying a short field pulse are of fundamental
interest. A short field pulse can be used to switch the core of an
antivortex.  This occurs through the rapid creation of a new 
antivortex and the annihilation of the original one with a short-lived
vortex. Second, we find that the static topological
complementarity of 
vortices and antivortices is equally exhibited in their dynamical
behavior. The dynamics of antivortices deserves further investigation
since it likely plays an equally important role as the dynamics of
vortices in the case of pulse-induced cross-tie domain-wall 
transformations~\cite{Neudert07}. In this context it will be necessary
to determine whether the pair-creation reported in
Ref.~\cite{Neudert07} is directly due to the dynamics described above,
since larger distances of a few hundreds of nm are involved in
cross-tie walls. Third, the annihilation of transient magnetic
structures during the antivortex core switching process could be used,
{\em e.g.}, for generating spin-waves for logical circuits
\cite{Hertel04}.

\end{document}